\documentclass{JHEP3}
\usepackage{amsmath}
\usepackage{amsthm}
\usepackage{amsfonts}
\usepackage{amssymb}

\def \Fig#1#2#3 {
\begin{figure}
\centering
\epsfxsize=#2cm \epsfbox{#1.eps}
\caption{#3}
\label{#1}
\end{figure}
}

\newcount\figno
\def\fig#1#2#3{
\par\begingroup\parindent=0pt\leftskip=1cm\rightskip=1cm\parindent=0pt
\baselineskip=15pt
\global\advance\figno by 1
\epsfxsize=#3
\centerline{\epsfbox{#2}}
\vskip 12pt
{\bf \small Figure \the\figno:} {\small #1}\par
\endgroup\par
}
\def\figlabel#1{\xdef#1{\the\figno
\mbox{ }}}
\def\encadremath#1{\vbox{\hrule\hbox{\vrule\kern8pt\vbox{\kern8pt
\hbox{$\displaystyle #1$}\kern8pt}
\kern8pt\vrule}\hrule}}

\title{ String cosmology from Poisson-Lie T-dual sigma models on supermanifolds }
\author{A. Eghbali and A. Rezaei-Aghdam\\
Department of Physics, Faculty of science, Azarbaijan
University of Tarbiat Moallem, 53714-161, Tabriz, Iran\\
E-mail: \email{a.eghbali@azaruniv.edu}, \email{rezaei-a@azaruniv.edu}
}


\abstract{We generalize the formulation of Poisson-Lie T-dual
sigma models on manifolds to supermanifolds. In this respect, we
formulate 1+1 dimensional string cosmological models on the Lie
supergroup $\bf C^3$ and its dual $\bf {(A_{1,1}+
2A)^0}_{1,0,0}$, which are coupled to two fermionic fields. Then,
we solve the equations of motion of the models and show that
there is a essential singularity for the metric of the original
model and its dual. }

\begin{document}





\section{Introduction}

Two-dimensional sigma models with supermanifolds as target space
have recently received considerable attention, because of their
relations to superstring models. The first attempt in this
direction dates back to about 3 decades ago \cite{Henneaun},
where the flat space GS superstring action was reproduced as a
WZW type sigma model on the coset superspace ($\frac{D=10~
Poincare~ supergroup}{{SO(9,1)}}$). Then, this work is extended
to the curved background \cite{Tseytlin} and shown that type IIB
superstring on $AdS_5 \otimes S^5$ can be constructed from sigma
model on the coset superspace $\frac{SU(2,2|4)}{SO(4,1)\times
SO(5)}$. After then, superstring theory on $AdS_3 \otimes S^3$ is
related to WZW model on $PSU(1,1|2)$ \cite{Berkovits} and also
superstring theory on $AdS_2 \otimes S^2$ is related to sigma
model on supercoset $\frac{PSU(1,1|2)}{U(1)\times U(1)}$
\cite{Bershadsky}. There are also other works in this direction,
see for instance \cite{Sorokin}.

\smallskip

On the other hand, T-duality is the most important symmetries of
string theory \cite{Busher}. Furthermore, Poisson-Lie T-duality, a
generalization of T-duality, does not require existence of
isometry in the original target manifold (as in usual T-duality)
\cite{K.S1}. So, the studies of Poisson-Lie T-duality in  sigma
models on supermanifolds and  duality in superstring theories on
AdS backgrounds are interesting problems. In the previous paper
\cite{ER} we extended Poisson-Lie symmetry to sigma models on
supermanifolds and also constructed Poisson-Lie T-dual sigma
models on Lie supergroups. In this paper, we formulate
Poisson-Lie T-dual sigma models on supermanifolds as an extension
of the  work \cite{K.S1}. Then, using this formalism we construct
1+1 dimensional string cosmological models as one of the first examples of
string cosmological models that have a Poisson-Lie symmetry.

\smallskip

The paper is organized as follows. In section two, we generalize
the formulation of Poisson-Lie T-dual sigma models on manifolds
to supermanifolds.  In section three, as an example of
Poisson-Lie T-dual sigma models on  supermanifolds, we construct
these models by using Lie supergroup $\bf C^3$ and its dual $\bf
{(A_{1,1}+ 2A)^0}_{1,0,0}$ and choosing orbit superspace as
one-dimensional space with coordinate $\{y^{\alpha} \}\equiv
\{t\}$ as time. In this way, we obtain 1+1 dimensional string
cosmological models which are coupled to two fermionic fields. In
this respect we give the one-loop beta functions equations for a
general sigma model on supermanifold \cite{Bershadsky} and write
string effective action on supermanifold in the beginning of this
section. Then, we solve the one-loop beta functions equations for
the original model and its dual and show  that there are some
solutions which  have singular points. Then, by writing  of the
Kretschmann scalar invariant on supermanifolds we show that some
of these singularities are essential for the model and its dual.
For self containing of the paper we write some mathematical
properties of matrices and tensors on supermanifold that we need
in this paper, as appendix.

\section{Super Poisson-Lie T-dual sigma models on
supermanifolds}

In the previous work \cite{ER}, we extended Poisson-Lie symmetry
to sigma models on supermanifolds and also constructed Poisson-Lie
T-dual sigma models on Lie supergroups. In this section as a
continuation of that work we formulate Poisson-Lie T-dual sigma
models on supermanifolds\footnote{ Here we generalize
 the results of Ref. \cite{K.S1} to the supermanifolds and in this direction we use the
 DeWitt notations \cite{D} for supermanifolds.}.
Consider a two-dimensional sigma model for the $d$ field
variables $X^{M}= (x^{\mu} , y^{\alpha})$ where $x^{\mu}$
($\mu=1,\cdots$,dim $G$) are coordinates of Lie supergroup $G$
that act freely from right on supermanifolds $M$. The
$\{y^{\alpha}\}$ are coordinates of the orbit $O=\frac{M}{G}$. In
this respect, one can construct a sigma model on $M$ with super
Poisson-Lie symmetry similar to \cite{K.S1} for ordinary
Poisson-Lie symmetry.

\smallskip

Consider a linear idempotent  map ${\cal{K}}(y)$: $T_y^{\ast} O
\oplus T_y O \oplus {\cal{D}} \longrightarrow T_y^{\ast} O \oplus
T_y O \oplus {\cal{D}}$ \cite{K.S1} where ${\cal{D}}$= ${\bf
\mathcal{G}}\oplus \tilde {\bf \mathcal{G}}$, \footnote {Let
$({\bf \mathcal{G}} , \tilde {\mathcal{G}} )$ be a Lie
superbialgebra \cite{An}, \cite{ER1}. There exists a unique Lie
superalgebra structures with the following commutation relations
on the vector space ${\bf \mathcal{G}}\oplus{\bf
\tilde{\mathcal{G}}}$ such that $\bf \mathcal{G}$ and $\tilde
{\mathcal{G}}$ are Lie superalgebras and the natural scalar
product on ${\bf \mathcal{G}}\oplus{\bf \tilde {\mathcal{G}}}$ is
invariant \cite{An}, \cite{ER4}
$$
[x , y]_{\cal{D}}\;=\;[x ,y],\;\; \;[x ,
\xi]_{\cal{D}}\;=\;-(-1)^{|x||\xi|}
ad^{\ast}_{\hspace{0.5mm}\xi}x+ad^{\ast}_{\hspace{0.5mm}x}\xi,\;\;\;[\xi
, \eta]_{\cal{D}}\;=\;[\xi ,\eta]_{{\bf \tilde {\mathcal{G}}}}\;\;\;\;\forall
x,y\in {\bf \mathcal{G}};\;\;  \xi, \eta \in {\bf \tilde {\mathcal{G}}},
$$
where
$$
<ad_x y\;,\;\xi>\;=\;-(-1)^{|x||y|}<y\;,\;
ad^{\ast}_{\hspace{0.5mm}x}\xi>,\;\;\;\;<ad_{\xi} \eta \;,\;x>\;=\;-(-1)^{|\xi||\eta|}<\eta\;,\;
ad^{\ast}_{\hspace{0.5mm}\xi}x>.
$$
The Lie superalgebra  ${\cal{D}} = {\bf \mathcal{G}}\oplus{\bf
\tilde{\mathcal{G}}}$ (related to the Lie supergroup $D$) is
called {\em Drinfel'd superdouble}.} is the Lie superalgebra of
the Drinfeld superdouble group $D$ of $G$ (with Lie superalgebra
${\bf \mathcal{G}}$). It has two eigen superspaces
${\cal{R}}_{\pm}(y) \subset T_y^{\ast} O \oplus T_y O \oplus
{\cal{D}} $ with eigenvalues $\pm 1$; such that dim
${\cal{R}}_{\pm}(y)=$dim $G$ + dim $M$. These eigen superspaces
may be considered as a graph of nondegenerate linear map
$E^{\pm}(y)$: $ T_y O \oplus  {\bf {\mathcal{G}}} \longrightarrow
T_y^{\ast} O \oplus {\tilde  {\bf {\mathcal{G}}}}$ \cite{K.S1}
\begin{equation}
{\cal{R}}_{\pm}(y)\;=\; Span\{ t\pm E^{\pm}(y)(t , .), \;\;t\in
T_y O \oplus {\bf {\mathcal{G}}} \}, \label{2.1}
\end{equation}
such that with translation of this graph to the point ${g} \in G$
we have
\begin{equation}
g^{-1} {\cal{R}}_{\pm}(y) g\;=\; Span\{ X_A \pm E^{\pm}_{AB}(g,y)
{\tilde X}^B \},\label{2.2}
\end{equation}
where $\{X_A\}=\{ {\overrightarrow{\partial}_{\alpha} }
=\frac{{\overrightarrow{\partial}}}{\partial y^{\alpha}}\; ,\;
X_i\}$ and $\{{\tilde X}^A\}=\{ {\overrightarrow {dy}^{\alpha}}  ,
{\tilde X}^i \}$ are the basis for the superspaces $T_y O \oplus
{{\bf {\mathcal{G}}}}$ and $T_y^{\ast} O \oplus {\tilde  {\bf {\mathcal{G}}}}$,
respectively. The matrix $E^{\pm}_{AB}(g,y)$ is a $d$-dimensional
matrix as follows:
\begin{equation}
E^{\pm}_{AB}(g,y)\;=\;\left( \begin{array}{cc}
                     E^{\pm}_{ij}(g , y) & \Phi^{\pm}_{i\beta}(g,y)\\

                     \Phi^{\pm}_{\alpha j}(g,y) & \Phi_{\alpha \beta}(y)
                      \end{array} \right),\label{2.3}
\end{equation}
where the minus sign (-) stands for supertranspose, i.e.,
${E^{+}}^{st}_{ij}= E^{-}_{ij} = (-1)^{ij} E^{+}_{ji}$,
$\Phi^{-}_{\alpha i}= (-1)^{i\alpha} \Phi^{+}_{i \alpha }$. Now,
similar to \cite{ER} we can write vector superspaces
${\cal{R}}_{\pm}(y)$ as follows:
\begin{equation}
g^{-1} {\cal{R}}_{\pm}(y) g\;=\; Span\{g^{-1} {X_A}  g \pm
E^{\pm}_{AB}(e,y) g^{-1} {\tilde X}^B g \},\label{2.4}
\end{equation}
such that
\begin{equation}
E^{\pm}_{AB}(e,y)\;=\;\left( \begin{array}{cc}
                     E^{\pm}_{0\;ij}\tiny(e , y) & F^{\pm}_{i\beta}(e,y)\\

                     F^{\pm}_{\alpha j}(e,y) & F_{\alpha \beta}(y)
                      \end{array} \right),\label{2.5}
\end{equation}
where $e$ is the unit element of $G$ such that $E^{\pm}_{0\;ij}(e
,y)$, $F^{\pm}_{i\beta}(e,y)$ and $ F^{\pm}_{\alpha \beta}(y)$ are
subsupermatrices with elements as functions of $y$ with
$F^{\mp}_{i \alpha} = (-1)^{i \alpha}F^{\pm}_{\alpha i}$. Now
using the following relations:
\begin{equation}
g^{-1} {X_A} g\;=\; {A{\tiny (\tiny g)}_A}^{B}\; {_B}X \;=\;
(-1)^B {A(g)_A}^{B}\; X_B,\label{2.6}
\end{equation}
\begin{equation}
g^{-1} {\tilde X^B} g\;=\; {B(g)}^{BC} {_C}X
+{D(g)}^{B}_{\;\;C}\;{\tilde X}^C \;=\; (-1)^C {B(g)}^{BC} X_C
+{D(g)}^{B}_{\;\;C}\;{\tilde X}^C,\label{2.7}
\end{equation}
one can obtain an expression for the  background matrix
$E^{\pm}_{AB}$
\begin{equation}
E^{\pm}_{AB}(g , y)= {A(g)_A}^{C}\;{{_C\Big(A(g)+E^{\pm}(e,y)
B(g)\Big)}^{-1}}^{D} {_DE_{B}^{\pm}(e,y)},\label{2.8}
\end{equation}
where {\footnotesize \begin{equation}{\footnotesize
{A(g)_A}^{B}=\left(
\begin{array}{cc} a(g)_i^{\;\;j}&0\\

                  0&(-1)^{\alpha} {_\alpha}{\delta}^{\;\beta}
                      \end{array} \right),\;{B(g)}^{BC}=\left(
\begin{array}{cc} b(g)^{jk}&0\\

                  0&0
                      \end{array} \right),\;{D(g)}^{B}_{\;\;C}=\left(
\begin{array}{cc} d(g)^{j}_{\;\;k}&0\\

                  0&{^\beta}{\delta}_{\gamma}
                      \end{array} \right).}\label{2.9}
\end{equation}}
Now in the same way as  \cite{{K.S1},{ER}} and using the
following equation of motion on the Drinfel'd  superdouble $D$:
\begin{equation}
<{\overrightarrow{\partial}_{\pm}l l^{-1} + w + v\; ,\;
{\cal{R}}_{\pm}(y)  }>=0,\;\;\;\;\;\;l\in D,\label{2.10}
\end{equation}
where $w = w_{\alpha} {\overrightarrow {dy}^{\alpha}}$ and
$v^{(L,l)}={v^{(L,l)}}^{\alpha}\frac{{\overrightarrow{\partial}}}{\partial
y^{\alpha}}$ are left invariant one-forms and left invariant
vector fields with left derivatives on the supercoset
$\frac{M}{G}$, respectively; one can obtain the following action
\begin{align}\label{2.11} \nonumber
S & = \  \frac{1}{2} \int  d\xi^{+} \wedge d\xi^{-} \Big[ {R_{+}^{(l)}}^{A}\; {_AE_{B}^{+}}(g,y)
\;{R_{-}^{(l)}}^{B}-\frac{1}{4} R^{(2)} { \varphi} \Big] \nonumber \\[2mm]
& = \ \frac{1}{2} \int d\xi^{+} \wedge d\xi^{-} \Big[ {R_{+}^{(l)}}^{i}\; {_iE_{j}^{+}}(g , y)
\;{R_{-}^{(l)}}^{j}+{R_{+}^{(l)}}^{i}\;
{_i{\Phi}_{\alpha}^{+}}(g,y) \;{\partial_{-}
}{y^{\alpha}} \nonumber  \\[2mm]
& + \ {\partial_{+}
}{y^{\alpha}}{_{\alpha}{\Phi}_{i}^{+}}(g,y)\;{R_{-}^{(l)}}^{i}
+{\partial_{+}
}{y^{\alpha}}{_{\alpha}{\Phi}_{\beta}}(y)\;{\partial_{-}
}{y^{\beta}} -\frac{1}{4} R^{(2)} { \varphi(g , y)  } \Big],\ \
\end{align}
where $R^{(2)}$ is the curvature of the world-sheet, $\varphi(g , y)$ is the dilaton field and
\begin{equation}
{R_{\pm}^{(l)}}^{A}=\{{R_{\pm}^{(l)}}^{i}\;,\;{\partial_{\pm}
}{y^{\alpha}}\},\;\;\;\;\;\;\;
{R_{\pm}^{(l)}}^{i}=({\partial}_{\pm}g\;g^{-1})^{i}={\partial}_{\pm}x^{\mu}\;
{{_\mu R}^{(l)}}^{i},\label{2.12}
\end{equation}
are right invariant one-forms with left derivatives. Furthermore,
using Eqs. \eqref{2.3}, \eqref{2.5} and \eqref{2.9} we have
\begin{align}\label{2.13} \nonumber
{{_iE_j}^{\pm}}(g , y) &  = \ {{{_i\Big((E_0^{\pm})^{-1}\pm
\Pi\Big)}_j}^{\hspace{-2mm}-1}}, \nonumber  \\[2mm]
{\Pi^{AB}}(g)  &  = \ {{B}^{AC}}(g)
\;{{_CA^{-1}}^B}(g),\;\;{\Pi^{ij}}(g)={{b}^{ik}}(g)
\;{{_ka^{-1}}^j}(g), \nonumber \\[2mm]
{{_i{\Phi}_{\alpha}}^{\pm}} &  = \ {{_iE_j}^{\hspace{-1mm}\pm}}\;{^j\Big((E_0^{\pm})^{-1}}\Big)^{k}
{{_k{F}_{\alpha}}^{\hspace{-1mm}\pm}},  \\[2mm]
{{_\alpha{\Phi}_{\beta}}} &  = \  {{_\alpha
F_\beta}}-{{_\alpha{F}_k}^{\hspace{-1mm}\pm}}\; \Pi^{kl}(g)\;
{{_lE_m}^{\hspace{-2mm}\pm}}(g)\;
{^m\Big((E_0^{\pm})^{-1}}\Big)^{n}{{_n{F}_\beta}^{\hspace{-1mm}\pm}}, \nonumber \\[2mm]
\varphi (g , y) &  = \  \varphi^0 (y^\alpha)+\ln sdet({{_iE_j}^{+}}(g , y))-\ln sdet({{_i{E_0^{+}}_{j}}}(e , y)) \nonumber.\ \
\end{align}
Note that the last equation is a quantum effect and it is a generalization of bosonic case \cite{Tyurin}.
In that equation, $\varphi^0$ is a the scalar field and function of the variable $y^{\alpha}$  only.
In the same way, one can obtain the dual sigma model (as \cite{Tyurin}, \cite{Sfetsos} and \cite{Von Unge} for the bosonic case)
\begin{align}\label{2.14} \nonumber
{\tilde S} & = \ \frac{1}{2} \int d\xi^{+} \wedge
d\xi^{-} \Big[ {{{\tilde R} _{+ A }}^{(l)}}\;{\tilde
E}^{+AB}(\tilde g,y) \;{_B{\tilde R}}_{-}^{(l)} -\frac{1}{4} R^{(2)} {\tilde \varphi}({\tilde g} , y) \Big]  \nonumber \\[2mm]
&  = \ \frac{1}{2} \int d\xi^{+} \wedge
d\xi^{-} \Big[ {{{\tilde R} _{+ i }}^{(l)}}\;{\tilde
E}^{+ij}(\tilde g , y) \;{_j{\tilde R}}_{-}^{(l)}+{{{\tilde R} _{+ i
}}^{(l)}}\;{^i\tilde \Phi}^{+}_{\; \alpha}(\tilde
g,y)\;{\partial_{-} }{y^{\alpha}}  \nonumber  \\[2mm]
&  + \ {\partial_{+}
}{y^{\alpha}}{_{\alpha}{\tilde \Phi}^{+\;i}}(\tilde
g,y)\;{_i{\tilde R}}_{-}^{(l)} +{\partial_{+}
}{y^{\alpha}}{_{\alpha}{\tilde \Phi}_{\beta}}(y)\;{\partial_{-}
}{y^{\beta}} -\frac{1}{4} R^{(2)} {\tilde \varphi(\tilde g , y)} \Big], \ \
\end{align}
where
\begin{equation}
{\tilde E}^{+AB}(\tilde g , y)= {{\Big(({{\tilde
E}^{+}}){^{-1}}(\tilde e,y)+{\tilde \Pi}(\tilde g ,
y)\Big)}^{-1}}^{AB},\label{2.15}
\end{equation}
such that
\begin{equation}
{\tilde E}^{+}(\tilde e, y)= {(A+{E}^{+}(e , y) B
)}^{-1}(C+{E}^{+}(e , y) D),\label{2.16}
\end{equation}
and
\begin{equation}
{\small A=\left(
\begin{array}{cc} 0&0\\

                  0& {_\alpha}{\delta}^{\;\beta}
                      \end{array} \right),\;\;B=\left(
\begin{array}{cc} {^i}{\delta}_{j}&0\\

                  0&0
                      \end{array} \right),\;\;C=\left(
\begin{array}{cc} {_i}{\delta}^{\;j}&0\\

                  0&0
                      \end{array} \right),\;\;D=\left(
\begin{array}{cc} 0&0\\

                  0&{^\alpha}{\delta}_{\beta}
                      \end{array} \right).}\label{2.17}
\end{equation}
Using Eq. \eqref{2.5} for ${\tilde E}^{AB}(\tilde e,y)$ one finds
\begin{equation}
{{\tilde E}_{0}^{+}}{^{ij}}(\tilde e) = {\Big((
E_{0}^{+})^{-1}}\Big){^{ij}}(e),~~~~~~~~~~\;\;\;\;\;\;\;\;{^i}{{\tilde
F}^{+}}_{\;
\alpha}={^i\Big((E_{0}^{+})^{-1}}\Big){^{j}}\;{_j}{{F}_{\alpha}}^{\hspace{-1mm}+},~~~~\label{2.18}
\end{equation}
\begin{equation}
{_\alpha}{{\tilde F}^{+}}{^{\;
i}}=-{_\alpha}{{F}_{j}}^{\hspace{-1mm}+}\;{^j\Big((E_{0}^{+})^{-1}}\Big){^{i}},\;\;\;\;\;{_\alpha}{{\tilde
F}}_{\; \beta}={_\alpha}{ F}_{\; \beta}-{_\alpha}\Big( F^{+}{(
E_{0}^{+})^{-1}}F^{+} \Big)_{\beta},\label{2.19}
\end{equation}
and from Eq. \eqref{2.15} we have
\begin{align}\label{2.20} \nonumber
{\tilde E}^{\pm ij}(\tilde g , y) &=\ {{\Big({{
E}_{0}}^{\pm}(e , y)+{\tilde \Pi}(\tilde g
)\Big)}^{-1}}^{ij},\\[2mm]
{^i}{{\tilde \Phi}^{\pm}}_{\;
\alpha} &=\ \pm {\tilde E}^{\pm ij} (\tilde g , y)
\;{_j}{{F}_{\alpha}}^{\hspace{-1mm}\pm}, \\[2mm]
{_\alpha}{{\tilde \Phi}}_{\; \beta} &=\ {_\alpha}{ F}_{\;
\beta}-{_\alpha} F^{\pm}{\tilde E^{\pm}}{^{kl}}(\tilde g , y)
\;{_l}{{F^{\pm}}}_{\; \beta},\nonumber \\[2mm]
{\tilde \varphi (\tilde g , y)} &=\ \varphi^0(y^\alpha)+ \ln sdet ({\tilde E}^{+ ij}(\tilde g , y)).\nonumber \ \
\end{align}
Note that in the above calculations,  all relations are the same as
\cite{K.S1} and \cite{Sfetsos}, but one must be careful that
matrices are supermatrices and in calculating their inverses,
products, etc, one must use from the rules of superinverses, superproducts and etc [see, appendix A].

\section{String cosmological  models}

Now by use of the actions  \eqref{2.11} and \eqref{2.14} one can
construct string cosmological models on supermanifold. For this
purpose we note that those actions in general, have the following
form:
\begin{equation}
S=\frac{1}{2} \int   d\xi^{+} \wedge
d\xi^{-} \Big[{\partial_{+}x}^{A}\; ({_AG}_{B}+
{_AB_{B}}){\partial_{-}x}^{B}-\frac{1}{4} R^{(2)} \varphi \Big].\label{3.7}
\end{equation}
The one-loop beta functions relations for the above sigma models
on supermanifolds have the following form \cite{Bershadsky}:
\begin{align}
{\beta^{(G)}}_{MN} &=\ R_{MN}+\frac{1}{4} H_{MPQ}
H^{QP}_{\;\;~~N}+2\overrightarrow{\nabla}_M \overrightarrow{\nabla}_N \varphi=0,\label{3.8.a} \\[2mm]
{\beta^{(B)}}_{NP} & =\ (-1)^M  \overrightarrow{\nabla}^M ( e^{-2\varphi} H_{MNP}) =0,\label{3.9} \\[2mm]
{{\beta}^{(\varphi)}} & =\ -R -\frac{1}{12} H_{MNP}H^{PNM}+4 \overrightarrow{\nabla}_M \varphi \overrightarrow{\nabla}^M \varphi  -4 \overrightarrow{\nabla}_M \overrightarrow{\nabla}^M \varphi=0,\label{3.10}~
\end{align}
where
\begin{align}
H_{MNP} \;=\; \frac{\overrightarrow{\partial}}{\partial
x^M}B_{NP} +(-1)^{M(P+N)}\frac{\overrightarrow{\partial}}{\partial
x^N}B_{PM}+(-1)^{P(M+N)}\frac{\overrightarrow{\partial}}{\partial
x^P}B_{MN},\label{3.11}
\end{align}
is the torsion field with the following  symmetry properties:
\begin{align}
H_{MNP} =  -(-1)^{PN} H_{MPN}\;=\; (-1)^{M(N+P)} H_{NPM}\;=\;
(-1)^{P(M+N)}H_{PMN}.
\end{align}
Indeed  the above beta functions relations are equations of motion
for the following effective action on supermanifold
\begin{align}
S_{eff}\;=\;\int d^{m, n}x\; \sqrt{G}e^{-2\varphi} [ R+4{\overrightarrow{\nabla}_M} {\varphi} \overrightarrow{\nabla}^M {\varphi}
+\frac{1}{12} H_{MNP}H^{PNM}],
\end{align}
where  $\sqrt{G}$ (with $G = sdet (_AG_B)$) and $d^{m, n}x$ are
measure and volume element on supermanifold, respectively,  with
$m$ bosonic and $n$ fermionic coordinates.  Furthermore, these
one-loop beta functions relations are Einstein field equations
which have coupled to bosonic and fermionic matters. For the bosonic
case, in Ref. \cite{N.M}, it has shown that the effective
action is invariant under Poisson-Lie T-duality; furthermore, it has obtained a
functional relation between one-loop beta functions of the
original and dual models and consequently showed that the
conformality of the models are invariant under Poisson-Lie
T-duality\footnote{Note that in those work it has shown that for
preserving of conformal invariant, the trace of the adjoint
representation of the structure constants related to  the Lie
group $G$ and its dual $\tilde G$ must be zero.}. In this way,
similar to the consequence of the previous section, we expect
that these proofs can be extended and satisfied to the case that
the target space is a supermanifold\footnote{Of course similar to
the bosonic case \cite{N.M} these proofs are very lengthy and we
leave those calculations to the another work.}.

\subsection{An example}

In this subsection we construct 1+1 dimensional string
cosmological models that are coupled to two fermionic fields. In
this respect, consider the Lie supergroup $\bf C^3$ with the
following Lie superalgebraic relation \cite{ER1}, \cite{B}:
\begin{equation}
[X_1\;,\;X_3]=X_2,\label{3.1}
\end{equation}
where its dual $\tilde {\bf \mathcal{G}} =(A_{1,1}+2A)^{0}_{1,0,0}$ has the
following anticommutation relation \cite{ER4}:
\begin{equation}
\{{\tilde X}^2\;,\;{\tilde X}^2\}={\tilde X}^1,\label{3.2}
\end{equation}
and nonzero (anti)commutation relations for the Drinfel'd superdouble ${\cal{D}}\;=\;(C^3 , (A_{1,1}+2A)^{0}_{1,0,0})$ have the following form\footnote{Note that these (anti)commutation relations have been written in the nonstandard basis. If  someone wishes to write these relations in the standard basis, it suffices
to multiply the structure constants of the anticommutators  by $i=\sqrt{-1}$.
} \cite{ER4}
$$
[X_1\;,\;X_3]=X_2,~~~~~~~~~~~~~~~~  \{{\tilde X}^2\;,\;{\tilde X}^2\}={\tilde X}^1,
$$
\vspace{-6mm}
\begin{equation}
[X_1\;,\;{\tilde X}^2]=-X_2-{\tilde X}^3,~~~~~~~~~~~~  \{ X_3 \;,\;{\tilde X}^2 \}=-{\tilde X}^1,
\end{equation}
such that the  $\{X_1 ,\tilde X^1\}$  and $\{ X_2 , X_3 , \tilde
X^2 , \tilde X^3 \}$  are bosonic and fermionic basis,
respectively. Note that as above discussion, the reason for choosing Lie supergroup
$\bf C^3$ is the fact that its adjoint representation is
traceless so that in this way the conformality under duality is
preserved \cite{N.M}. Now, choosing parametrization for the Lie
supergroups $\bf C^3$ and $\bf (A_{1,1}+2A)^{0}_{1,0,0}$ we
construct the model and its dual.

\vspace{5mm}

\subsubsection{  Model}

\vspace{3mm}

For the Lie supergroups $\bf C^3$ we choose the following
parametrization:\footnote{Note that the structure of supergroup
is fully defined by specifying a coproduct and antipode on the
space of functions on it. Our local parametrizations \eqref{3.3}
and \eqref{3.34} together with the superalgebra relations
\eqref{3.1} and \eqref{3.2} give however an equivalent
information if we take the respective underlying bosonic
subgroups contractible.  }
\begin{equation}
g=e^{xX_1} e^{\psi X_2} e^{\chi X_3},\label{3.3}
\end{equation}
where the  $x$ is bosonic parameter and $\psi, \chi$ are fermionic
ones. Now using Eqs. \eqref{2.6}, \eqref{2.7} and \eqref{2.13} we
have
\begin{equation}
\Pi^{ij} (g)=\left( \begin{array}{ccc}
                    0 & 0 & 0\\
                    0 & -x & 0\\
                    0 &0 & 0
                      \end{array} \right).\label{3.4}
\end{equation}
Finally, choosing the orbit $O=\frac{M}{G}$ as a one-dimensional
space with time coordinate $\{y^{\alpha}  \}\equiv \{t\}$, using
the Eqs. \eqref{2.11},  \eqref{2.13} and assuming  ${_\alpha
F}_{\beta}=f(t)$ and ${_i {F^{\pm}}}_{\beta}=0$ we obtain the
following action for the original model:
\begin{align}
S &=\ \frac{1}{2} \int d\xi^{+} \wedge
d\xi^{-}
\Big[ {\partial_{+}}t~f(t)~ {\partial_{-}}t+{\partial_{+}}x
~\frac{1}{a(t)}~ {\partial_{-}}x-{\partial_{+}}\psi~
\frac{1}{e(t)}~{\partial_{-}}\chi \nonumber \\[2mm]
&+\ {\partial_{+}}\chi
~\frac{1}{e(t)}~{\partial_{-}}\psi-{\partial_{+}}\chi
~\frac{x}{e^2(t)}~{\partial_{-}}\chi -\frac{1}{4} R^{(2)} \varphi^0(t)\Big],\label{3.5}
\end{align}
where we have chosen  the constant matrix $({E_0}^{+})^{-1}$ as
follows:
\begin{equation}
{( E_{0}^{+})^{-1}}{^{ij}}=\left( \begin{array}{ccc}
                    a(t) & 0 & 0\\
                    0 & 0 & e(t)\\
                    0 & -e(t) & 0
                      \end{array} \right).\label{3.6}
\end{equation}
Such that for this model we have
\begin{equation}
{G}_{AB}=\left( \begin{array}{cccc}
                    f(t) & 0 & 0 & 0\\
                    0 & \frac{1}{a(t)} & 0& 0\\
                    0 & 0 & 0& \frac{1}{e(t)}\\
                    0 & 0 & -
                    \frac{1}{e(t)}&0
                      \end{array} \right),~~~~~~~~~~{B}_{AB}=\left( \begin{array}{cccc}
                    0 & 0 & 0 & 0\\
                    0 & 0 & 0& 0\\
                    0 & 0 & 0& 0\\
                    0 & 0 & 0&\frac{x}{e^2(t)}
                      \end{array} \right).
\end{equation}
Now one can construct the beta functions equations
\eqref{3.8.a}-\eqref{3.10} for the action \eqref{3.5} with
assuming  $\varphi = \varphi^{0} (t)=0$. For this action using
\eqref{3.11} one can obtain the  nonzero components of $H_{MNP}$
as follows:
\begin{equation}
H_{033}\;=\;x \frac{d}{dt} (\frac{1}{e^2(t)}),~~~~~~~~~~~~ H_{133}\;=\;(\frac{1}{e^2(t)}),\label{3.12.a}
\end{equation}
so for this example $H_{MN} = H_{MPQ} H^{QP}_{\;\;~~N} = 0$ and
$H^{2} = H_{MNP}H^{PNM} = 0$; hence, the beta function relations
are rewritten as follows:
\begin{align}
{\beta^{(G)}}_{MN} &=\ R_{MN}=0,\label{3.12}\\[2mm]
{\beta^{(B)}}_{NP} &=\ (-1)^M {\overrightarrow{\nabla}^M}
H_{MNP}=0,\label{3.12.b}\\[2mm]
{\beta}^{(\varphi)}&=\  R=0,\label{3.13}
\end{align}
where the nonzero components of $R_{MN}$ are $R_{00}, R_{11}$ and $R_{23}$.
Note that in this way we have a Ricci flat supermanifold. After some calculations the relations \eqref{3.12} have the
following forms
\begin{align}
R_{00} &=\ \frac{1}{2}\Big
[\frac{d}{dt}(\frac{\dot{a}}{a})-\frac{\dot{a}\dot{f}}{2af}-\frac{\dot{a}^2}{2a^2}-2\frac{d}{dt}(\frac{\dot{e}}{e})
+\frac{\dot{e}\dot{f}}{ef}+\frac{\dot{e}^2}{e^2}\Big]=0,\label{3.14}
\\[2mm]
R_{11} &=\  \frac{1}{2}\Big
[\frac{d}{dt}(\frac{\dot{a}}{a^2f})+\frac{\dot{a}\dot{f}}{2a^2f^2}+\frac{\dot{a}^2}{2a^3f}
+\frac{\dot{e}\dot{a}}{a^2ef}\Big]=0,\label{3.15} \\[2mm]
R_{23} &=\ \frac{1}{2}\Big
[\frac{d}{dt}(\frac{\dot{e}}{e^2f})+\frac{\dot{e}\dot{f}}{2e^2f^2}-\frac{\dot{e}\dot{a}}{2e^2af}
+2\frac{\dot{e}^2}{e^3f}\Big]=0,\label{3.16}
\end{align}
where dot stands for time derivative. Furthermore, the relation
\eqref{3.13} leads to
\begin{equation}
R = \frac{1}{f}\Big [\frac{d}{dt}(\frac{\dot{a}}{a}-2\frac{\dot{e}}{e})+\frac{\dot{e}\dot{f}}{ef}
-\frac{\dot{a}\dot{f}}{2af}+\frac{\dot{a}\dot{e}}{ae}-\frac{\dot{e}^2}{2e^2}
-\frac{\dot{a}^2}{2a^2}\Big]=0,\label{3.17}
\end{equation}
and only the nonzero component of ${\beta^{(B)}}_{NP}$ is
\begin{equation}
{\beta^{(B)}}_{33}=  \frac{-2x}{fe^2}\Big
[\frac{d}{dt}(\frac{\dot{e}}{e})-\frac{\dot{e}\dot{f}}{2ef}-\frac{\dot{e}\dot{a}}{2ea}
+\frac{\dot{e}^2}{e^2}\Big]=
\frac{-4x}{e(t)}R_{23}=0.\label{3.17.a}
\end{equation}
Now, by combination of Eqs. \eqref{3.15}, \eqref{3.16} and
\eqref{3.17} we obtain the following constraints:
\begin{equation}
\frac{d}{dt}\ln(e(t))=0,\label{3.18}~~~~~~~~~~~~~~
\end{equation}
\begin{equation}
\frac{d}{dt}\ln(a(t))=\frac{3}{2}\frac{d}{dt}\ln(e(t)).\label{3.19}
\end{equation}
After substituting  the constraint \eqref{3.18} into the Eq.
\eqref{3.14} we obtain the following equation:
\begin{equation}
\frac{d^2}{dt^2}\ln(a(t))-\frac{1}{2}\frac{d}{dt}\ln(a(t))\frac{d}{dt}\ln(f(t))
-\frac{1}{2}\Big(\frac{d}{dt}\ln(a(t))\Big)^2=0.\label{3.20}
\end{equation}
The general solution for the above equation has the following
form:
\begin{equation}
a(t)\;=\;c_1 e^{-2A(t)}, \label{3.20.a}
\end{equation}
where
\begin{equation}
A(t)\;=\;{{\int^t \frac{\sqrt{f(t')}dt'}{[\int^{t'}
\sqrt{f(t'')}dt''+c_2]}}}~~~~~~~~~~c_1, c_2 \in \Re.
\label{3.20.b}
\end{equation}
Furthermore one can obtain the following special class of
solutions for the above equation:
\begin{equation}
(i)\;\;\;\; a(t)=\frac{a_0}{ (t-\alpha_0)},\;\;\;\;\;e(t)=e_0
,\;\;\;\;\;f(t)=\frac{f_0 }{(t-\alpha_0)},\hspace{3cm}\label{3.21}
\end{equation}
\vspace{-5mm}
\begin{equation}
(ii)\;\;\;a(t)=a_0 e^{b_0t},\;\;\;\;\;e(t)=e_0
,\;\;\;\;\;f(t)=f_0 e^{-b_0t}.\hspace{4cm}\label{3.22}
\end{equation}
On the other hand, by substituting the constraint \eqref{3.19}
into the Eq. \eqref{3.14} we obtain the following equation:
\begin{equation}
\frac{d^2}{dt^2}\ln(e(t))-\frac{1}{2}\frac{d}{dt}\ln(f(t))\frac{d}{dt}\ln(e(t))+
\frac{1}{4}\Big(\frac{d}{dt}\ln(e(t))\Big)^2=0.\label{3.23}
\end{equation}
The general solution for the above equation has the following
form:
\begin{equation}
e(t)\;=\;c_1 e^{4A(t)}, \label{3.20.d}
\end{equation}
for which we obtain the following special class of solutions:
\begin{equation}
(iii)\;\;\;a(t)=a_0 e^{\frac{3}{2}c_0t},\;\;\;\;\;e(t)=e_0
e^{c_0t},\;\;\;\;\;f(t)=f_0
e^{\frac{1}{2}c_0t},\hspace{3cm}\label{3.24}
\end{equation}
\vspace{-6mm}
\begin{equation}
(iv)\;\;\;\;a(t)=\frac{a_0}
{(t-\beta_0)^3},\;\;\;\;\;e(t)=\frac{e_0}
{(t-\beta_0)^2},\;\;\;\;\;f(t)=\frac{f_0}{
(t-\beta_0)^3},~~~~~~~~~~~\label{3.25}
\end{equation}
\vspace{-2mm}
\begin{equation}
(v)\;\;\;\;a(t)=a_0 (t-\gamma_0)^3,\;\;\;\;\;e(t)=e_0
(t-\gamma_0)^2,\;\;\;\;\;f(t)=\frac{f_0}{
(t-\gamma_0)},~~~~~~~~\label{3.26}
\end{equation}
where $a_0, e_0, f_0, b_0, c_0, \alpha_0, \beta_0 $ and $\gamma_0
$ are real constants. We see that the class \eqref{3.21}, \eqref{3.25} and
\eqref{3.26} of solutions have singular points at $t=\alpha_0,
t=\beta_0$ and $t=\gamma_0$, respectively. To investigate the type
of singular points we write the Kretschmann scalar invariant for
supermanifold as follows:
\begin{equation}
{K}\;=\;R^{IJKL}R_{IJKL}.\label{3.27}
\end{equation}
Using the matrix representation, we rewrite this formula in the
following form\footnote{Note that this matrix representation  is
useful for simplifying the computations.}:
{\small $$ K
\;=\;-(-1)^{I+J+L+IJ+IN+IL+JL+MK+NK+LK} G^{KP}\;{_P(R_{MN})}_Q \;
G^{QL}\;{_L(R_{IJ})}_K\; G^{JN} G^{IM}\hspace{10mm}
$$}
\vspace{-3mm}
{\small \begin{equation}
=\;-(-1)^{I+J+K+L+IJ+IK+IL+JK+JL+IN+LK} {\bf str}
\Big(G^{-1}\;R_{MN} \; G^{-1}\;R_{IJ}\Big) G^{JN}
G^{IM},\label{3.28}
\end{equation}}
where $(R_{MN})_{PQ}\;=\;R_{MNPQ}$. Now using the following form
of the metric of the original model
$$
ds^2\;=\;dx^A\;_AG_B\; dx^B\;=\; (-1)^{AB}\; G_{AB} \;dx^A\; dx^B
\hspace{2cm}
$$
\vspace{-4mm}
\begin{equation}
= f(t)dt^2+\frac{1}{a(t)}\;dx^2-\frac{1}{e(t)}\;d\psi
d\chi+\frac{1}{e(t)}\;d\chi d\psi,~~~\label{3.29}
\end{equation}
and after some calculations one can obtain the general form of the
Kretschmann scalar invariant for the model as follows:
\begin{equation}
K = \frac{1}{f^2}\Big[\Big(
\frac{d}{dt}(\frac{\dot{a}}{a})-\frac{\dot{a}\dot{f}}{2af}-
\frac{\dot{a}^2}{2a^2} \Big)^2+2\Big(
\frac{d}{dt}(\frac{\dot{e}}{e})-\frac{\dot{e}\dot{f}}{2ef}-
\frac{\dot{e}^2}{2e^2} \Big)^2 +\frac{\dot{a}^2 \dot{e}^2}{2a^2
e^2}+\frac{3\dot{e}^4}{4e^4} \Big].\label{3.30}
\end{equation}
For solutions $(i)$ and $(ii)$ the Kretschmann scalar invariant
vanishes and for the solutions  $(iii), (iv)$ and $(v)$  we have
\begin{equation}
K_{(iii)} = \frac{21 {c_0}^4}{4{f_0}^2}\;e^{-c_0 t},~~\label{3.31}
\end{equation}
\begin{equation}
K_{(iv)} = \frac{84}{{f_0}^2} \;(t-\beta_0)^2,\label{3.32}
\end{equation}
\begin{equation}
K_{(v)} = \frac{84}{{f_0}^2\;(t-\gamma_0)^2}.\label{3.33}
\end{equation}
We see that in the latter case the Kretschmann scalar invariant is
singular for the point $t=\gamma_0$; therefore this  singular
point is {\it essential}.

\vspace{4mm}
\subsubsection{Dual model}

\vspace{3mm}

In the same way, one can construct the dual model on the Lie
supergroup $\bf (A_{1,1}+2A)^{0}_{1,0,0}$ using the following
parametrization:
\begin{equation}
\tilde g=e^{\tilde x \tilde X^1} e^{\tilde \psi \tilde X^2}
e^{\tilde \chi \tilde X^3}.\label{3.34}
\end{equation}
In this case, we find
\begin{equation}
{\tilde \Pi}_{ij}(\tilde g)=\left( \begin{array}{ccc}
                    0 & 0 & -\tilde \psi\\
                    0 & 0 & 0\\
                    \tilde \psi &0 & 0
                      \end{array} \right),\label{3.35}
\end{equation}
and using the equations \eqref{2.14}-\eqref{2.20} the dual action
is obtained as
\begin{align}\nonumber
\tilde S &=\ \frac{1}{2} \int  d\xi^{+} \wedge
d\xi^{-} \Big[ {\partial_{+}}t \;f(t)\;
{\partial_{-}}t+{\partial_{+}}{\tilde x}\; {a(t)}\;
{\partial_{-}}{\tilde x}+{\partial_{+}}{\tilde
x}\;\Big(a(t)e(t)\tilde \psi-\frac{a(t)\tilde
\psi}{2}\Big)\;{\partial_{-}}{\tilde \psi}\\[2mm]
&+\ {\partial_{+}}{\tilde \psi}\;\Big(a(t)e(t)\tilde
\psi+\frac{a(t)\tilde \psi}{2}\Big)\;{\partial_{-}}{\tilde x}-
{\partial_{+}}{\tilde \psi}\;e(t)\;{\partial_{-}}{\tilde
\chi}+{\partial_{+}}{\tilde \chi}\;e(t)\;{\partial_{-}}{\tilde
\psi} -\frac{1}{4}R^{(2)} {\tilde \varphi }\Big],\label{3.36}
\end{align}
such that for this model we have
\begin{equation}
{\tilde G}_{AB}=\left( \begin{array}{cccc}
                    f(t) & 0 & 0 & 0\\
                    0 & a(t) & -\frac{a(t) \tilde \psi}{2}& 0\\
                    0 & -\frac{a(t) \tilde \psi}{2} & 0& {e(t)}\\
                    0 & 0 & -
                    {e(t)}&0
                      \end{array} \right),~~~~~~~~~~{\tilde B}_{AB}=\left( \begin{array}{cccc}
                    0 & 0 & 0 & 0\\
                    0 & 0 & a(t)e(t)\tilde \psi& 0\\
                    0 & -a(t)e(t)\tilde \psi & 0& 0\\
                    0 & 0 & 0&0
                      \end{array} \right).
\end{equation}
Note that  for the above action the nonzero components of $\tilde H$ have
the following forms:
\begin{equation}
\tilde H_{012}~=~\frac{d}{dt}\Big(a(t)e(t)\Big) \tilde
\psi,~~~~~~~~~~\tilde H_{122} = -2 a(t)e(t).\label{3.37}
\end{equation}
Also, by taking the $\varphi =\varphi^0 (t) =0$ and using the last equation in \eqref{2.20} we find
\begin{equation}
{\tilde \varphi} = \ln (\frac{a(t)}{e^2 (t)}).\label{3.37.1}
\end{equation}
Now using the Eq. \eqref{3.37} we obtain that ${\tilde H}_{MN} =
{\tilde H}_{MPQ} {\tilde H}^{QP}_{\;\;~~N} = 0$ and ${\tilde
H}^{2} = {\tilde H}_{MNP}{\tilde H}^{PNM} \\= 0$; in this way the
relations \eqref{3.8.a}-\eqref{3.10} take the following forms for
the dual model:
\begin{align}
{\beta^{(\tilde G)}}_{MN} &=\ {\tilde R}_{MN}+2\overrightarrow{\nabla}_M \overrightarrow{\nabla}_N {\tilde \varphi}=0,\label{3.8} \\[2mm]
{\beta^{(\tilde B)}}_{NP} & =\ (-1)^M  \overrightarrow{\nabla}^M ( e^{-2\tilde \varphi}{\tilde  H}_{MNP}) =0,\label{3.9} \\[2mm]
{{\beta}^{(\tilde \varphi)}} & =\ -\tilde R +4(\overrightarrow{\nabla}{\tilde \varphi})^2 -4\nabla^2 {\tilde \varphi}=0,\label{3.37.2}~
\end{align}
where the nonzero components of ${\tilde R}_{MN}$ and $\overrightarrow{\nabla}_M \overrightarrow{\nabla}_N {\tilde \varphi}$ have the following forms, respectively, (see, appendix A)
\vspace{-2mm}
\begin{align}
{\tilde R}_{00} &=\ \frac{d}{dt}(\frac{\dot{e}}{e}- \frac{\dot{a}}{2a})+\frac{\dot{a}\dot{f}}{4af}-
\frac{\dot{e}\dot{f}}{2ef}-\frac{\dot{a}^2}{4a^2}+\frac{\dot{e}^2}{2e^2},  \label{3.37.a}\\[2mm]
{\tilde R}_{11} & =\ -\frac{d}{dt}(\frac{\dot{a}}{2f})-\frac{\dot{a}\dot{f}}{4f^2}+
\frac{\dot{a}\dot{e}}{2ef}+\frac{\dot{a}^2}{4af}, \\[2mm]
{\tilde R}_{12} & =\ \frac{-1}{2} {\tilde R}_{11} \tilde
\psi,\\[2mm]
{\tilde R}_{23} & =\
-\frac{d}{dt}(\frac{\dot{e}}{2f})-\frac{\dot{e}\dot{f}}{4f^2}-
\frac{\dot{a}\dot{e}}{4af}+\frac{\dot{e}^2}{ef},
\end{align}
\vspace{-5mm}
\begin{align}
\overrightarrow{\nabla}_0 \overrightarrow{\nabla}_0 {\tilde
\varphi} &=\ \frac{d^2}{dt^2} \ln (\frac{a}{e^2})-
\frac{\dot{f}}{2f} \frac{d}{dt} \ln (\frac{a}{e^2}),
~~~~~~~\overrightarrow{\nabla}_1 \overrightarrow{\nabla}_1
{\tilde \varphi}\;=\;\frac{\dot{a}}{2f}
\frac{d}{dt} \ln (\frac{a}{e^2}), \label{3.38.a}\\[2mm]
\overrightarrow{\nabla}_1 \overrightarrow{\nabla}_2 {\tilde \varphi} & =\
 -\frac{{1}}{2} (\overrightarrow{\nabla}_1 \overrightarrow{\nabla}_1 {\tilde \varphi}) \tilde
 \psi,~~~~~~~~~~~~~~~~~~~~
 \overrightarrow{\nabla}_2 \overrightarrow{\nabla}_3 {\tilde \varphi}\;=\; \frac{\dot{e}}{2f} \frac{d}{dt} \ln (\frac{a}{e^2}),
\end{align}
and the ${\beta^{(\tilde B)}}_{NP}=0$ leads to
\begin{align}
&(-2\frac{\overrightarrow{\partial}}{\partial {\tilde x}^M}{\tilde \varphi}){\tilde H}^M_{\;~NP} +(-1)^M \; \overrightarrow{\nabla}^M {\tilde H}_{MNP} \nonumber \\
&=\ (-2\frac{\overrightarrow{\partial}}{\partial {\tilde x}^M}{\tilde \varphi}){\tilde H}^M_{\;~NP} +(-1)^{M+L+L(N+P)}\;{\tilde G}^{LM}[ {\tilde H}_{MNP} \frac{\overleftarrow{\partial}}{\partial {\tilde x}^L} \nonumber \\[2mm]
&-\ (-1)^{(N+P)(M+Q)}\;{\tilde H}_{QNP} {\tilde \Gamma^Q}_{\;~ML}
-(-1)^{P(N+Q)} \; {\tilde H}_{MQP} {\tilde \Gamma^Q}_{\;~NL}-{\tilde H}_{MNQ} {\tilde \Gamma^Q}_{\;~PL} ]=0. \label{3.37.b}~
\end{align}
The dilatonic contribution to the ${\beta^{(\tilde \varphi)}}$ is
\begin{align}
(\overrightarrow{\nabla}{\tilde \varphi})^2 - \nabla^2 {\tilde \varphi}& =\  \; (\overrightarrow{\nabla}_M {\tilde \varphi})
(\overrightarrow{\nabla}^M {\tilde \varphi})- \; \overrightarrow{\nabla}_M \overrightarrow{\nabla}^M {\tilde \varphi} \nonumber\\[2mm]
&=\ (-1)^{M+N} {\tilde G}^{MN} \Big[({\tilde \varphi}\frac{\overleftarrow{\partial}}
{\partial {\tilde x}^N}) ({\tilde \varphi}\frac{\overleftarrow{\partial}}{\partial {\tilde x}^M})-
 ({\tilde \varphi}\frac{\overleftarrow{\partial}}{\partial {\tilde x}^N})\frac{\overleftarrow{\partial}}
 {\partial {\tilde x}^M}+  ({\tilde \varphi}\frac{\overleftarrow{\partial}}{\partial {\tilde x}^P})
  {\tilde \Gamma^P}_{\;~NM} \Big]\nonumber\\[2mm]
&=\ \frac{1}{f}\Big[ \frac{d}{dt} (
\frac{2\dot{e}}{e}-\frac{\dot{a}}{a})+
 \frac{\dot{a}^2}{2a^2}+\frac{2\dot{e}^2}{e^2}
 + \frac{\dot{a}\dot{f}}{2af}- \frac{\dot{e}\dot{f}}{ef}- \frac{2\dot{a}\dot{e}}{ae}\Big]. \label{3.37.4}~
\end{align}
Finally, by substituting the relations \eqref{3.37.a}-\eqref{3.37.4} into the Eqs. \eqref{3.8}-\eqref{3.37.2} we obtain the following equations:
\begin{equation}
\frac{d}{dt}(\frac{\dot{a}}{a}-\frac{2\dot{e}}{e})-\frac{\dot{a}\dot{f}}{2af}+
\frac{\dot{e}\dot{f}}{ef}-\frac{\dot{a}^2}{6a^2}+\frac{\dot{e}^2}{3e^2}=0,\;\;~~~~~~~~~~~~~\label{3.38}
\end{equation}
\vspace{-5mm}
\begin{equation}
\frac{d}{dt}(\frac{\dot{a}}{a})-\frac{\dot{a}\dot{f}}{2af}-\frac{3\dot{a}^2}{2a^2}
+\frac{3\dot{a}\dot{e}}{ae}=0,\;\;\label{3.39}
\end{equation}
\vspace{-3mm}
\begin{equation}
\frac{d}{dt}(\frac{\dot{e}}{e})-\frac{\dot{e}\dot{f}}{2ef}-\frac{3\dot{a}\dot{e}}{2ae}
+\frac{3\dot{e}^2}{e^2}=0,~\label{3.40}
\end{equation}
\vspace{-3mm}
\begin{equation}
\frac{d}{dt}(\frac{\dot{a}}{a}+\frac{\dot{e}}{e})
-\frac{\dot{a}\dot{f}}{2af}-\frac{\dot{e}\dot{f}}{2ef}+\frac{3\dot{a}\dot{e}}{2ae}-\frac{3\dot{a}^2}{2a^2}
+\frac{3\dot{e}^2}{e^2}=0,~~~~~~~~~~~~~~~~~~~~~~\label{3.41}
\end{equation}
\vspace{-2mm}
\begin{equation}
\frac{d}{dt}(\frac{\dot{a}}{a}-\frac{2\dot{e}}{e})
-\frac{\dot{a}\dot{f}}{2af}+\frac{\dot{e}\dot{f}}{ef}+\frac{3\dot{a}\dot{e}}{ae}-\frac{5\dot{a}^2}{6a^2}
-\frac{17\dot{e}^2}{6e^2}=0,\;~~~~~~~~~~~~~~~~~~~~\label{3.42}
\end{equation}
Now, by combination of Eqs. \eqref{3.38} and \eqref{3.42} we find
the following constraint:
\begin{equation}
\frac{d}{dt}\ln(a(t))=\frac{9 \pm \sqrt{5}
}{4}\frac{d}{dt}\ln(e(t)), \label{3.42.s}
\end{equation}
and the result of  combination  Eqs. \eqref{3.39} and
\eqref{3.41} is Eq. \eqref{3.40}, then by substituting the
constraint \eqref{3.42.s} into Eq. \eqref{3.40}, we obtain the
following equation:
\begin{equation}
\frac{d^2}{dt^2}\ln(e(t))-\frac{1}{2}\frac{d}{dt}\ln(e(t))\frac{d}{dt}\ln(f(t))-\frac{3(1\pm
\sqrt{5} )}{8} \Big(\frac{d}{dt}\ln(e(t))\Big)^2=0,\label{3.43}
\end{equation}
where the general solution for the above equation has the
following form:
\begin{equation}
e(t)\;=\;c_1 e^{{-\frac{8}{3(1\pm \sqrt{5} )} }A(t)},
\label{3.20.a}
\end{equation}
for which we obtain the following special class of solutions
\begin{equation}
(i)^{\pm}:\;\;\;\; a(t)={\tilde a}_0 e^{\frac{(9\pm
\sqrt{5})}{4}{\tilde \alpha}_0 t},\;\;\;\;\;e(t)={\tilde e}_0
e^{{\tilde \alpha}_0 t } ,\;\;\;\;\;f(t)={\tilde f}_0
e^{\frac{-3(1\pm \sqrt{5})}{4}{\tilde \alpha}_0
t},\hspace{1.5cm}\label{3.42.a}
\end{equation}
\vspace{-6mm}
\begin{equation}
(ii)^{\pm}:\;\;\;a(t)={\tilde a}_0 (t-{\tilde \beta }_0)^{\frac{9
\pm \sqrt{5}}{12}},\;\;\;\;\;e(t)={\tilde e}_0 (t-{\tilde \beta
}_0)^{\frac{1}{3}} ,\;\;\;\;\;f(t)=\frac{{\tilde f}_0}{
(t-{\tilde \beta }_0)^{\frac{9 \pm
\sqrt{5}}{4}}},~~~~\label{3.42.b}
\end{equation}
where ${\tilde a}_0, {\tilde e}_0, {\tilde f}_0, {\tilde
\alpha}_0$ and ${\tilde \beta}_0$ are real constants. Using the following
form of the metric of the dual model
\begin{equation}
d\tilde {s}^2\;= \;f(t)dt^2+{a(t)}\;d\tilde
{x}^2-\frac{a(t)\tilde \psi}{2}\;d\tilde {\psi} d\tilde
{x}-\frac{a(t)\tilde \psi}{2}\;d\tilde {x} d\tilde {\psi}
-{e(t)}\;\tilde {d\psi} \tilde {d\chi}+{e(t)}\;d\tilde {\chi}
d\tilde {\psi},~~\label{3.48}
\end{equation}
and after some calculations one can obtain the general form of the
Kretschmann scalar invariant for the dual model as follows:
\begin{equation}
\tilde K = \frac{1}{f^2}\Big[\Big(
\frac{d}{dt}(\frac{\dot{a}}{a})-\frac{\dot{a}\dot{f}}{2af}+
\frac{\dot{a}^2}{2a^2} \Big)^2+2\Big(
\frac{d}{dt}(\frac{\dot{e}}{e})-\frac{\dot{e}\dot{f}}{2ef}+
\frac{\dot{e}^2}{2e^2} \Big)^2 +\frac{\dot{a}^2 \dot{e}^2}{2a^2
e^2}+\frac{3\dot{e}^4}{4e^4} \Big].\label{3.49}
\end{equation}
For solutions $(i)^{\pm}$,  $(ii)^{+}$ and $(ii)^{-}$ the
Kretschmann scalar invariant is given by
\begin{align}
{\tilde K}_{(i)^{\pm}} &=\ \frac{(269 \pm 111\sqrt{5}){{\tilde
\alpha }_0}^4}{8{{\tilde f }_0}^2}\;e^{\frac{3}{2}(1\pm \sqrt{5}
){{\tilde \alpha }_0}t},\label{3.31.m} \\[2mm]
{\tilde K}_{(ii)^{+}} &=\ \frac{22.82}{{{\tilde f }_0}^2}
\;(t-{\tilde \beta}_0)^{\frac{1+\sqrt{5}}{2}},\label{3.32.n} \\[2mm]
{\tilde K}_{(ii)^{-}} &=\ \frac{2.28}{{{\tilde f }_0}^2}
\;\frac{1}{(t-{\tilde \beta}_0)^{\frac{\sqrt{5}-1}{2}}}.
\label{3.33.l}
\end{align}
We see that in the latter case the Kretschmann scalar invariant is
singular for the point $t =  {\tilde \beta}_0$; therefore this
singular point is essential. Note that the form and coefficients
of the Kretschmann scalar invariants for the original model and
its dual are the same for all solutions and as we expect the feature of essential singularity
of the metric  of the model and its dual are preserved under duality, because duality transformation
is a canonical transformation.


\section{Conclusion}

In this paper, as a continuation of Ref. \cite{ER} we extended the
results of Ref. \cite{K.S1} to the supermanifolds by using of the
formulation of Poisson-Lie T-dual sigma models
 on supermanifolds. Then, using this formalism we constructed 1+1 dimensional string cosmological models
 as an example which has super Poisson-Lie symmetry. Also one can construct other models by
 using other Lie superbialgebras\footnote{Note that these Lie superbialgebras have zero supertrace for the adjoint representation
 of the generators so that in this way the conformality is preserved under duality
  transformation \cite{N.M}.} such as
 $(C^5_{p=0} , {\tilde {\bf \mathcal{G}}}_{\alpha, \beta, \gamma})$ and
 $(C^2_{p=-1} , {\tilde {\bf \mathcal{G}}}_{\alpha, \beta, \gamma})$ of
 Ref. \cite{ER4}. Furthermore, in this way one can construct the 2+1 and
 3+1 dimensional string cosmological models that have super Poisson-Lie symmetry \cite{ER5}.

\bigskip
\bigskip

\noindent {\bf Acknowledgments:} This research was supported by a research fund No. 401.231 from Azarbaijan university
of Tarbiat Moallem. We would like to thank F. Darabi and M. Atazadeh for carefully reading the
manuscript and useful comments.


\bigskip
\appendix

\section{Some properties of matrices and tensors on supervector
space and supermanifolds }

In this appendix we collect a few relevant details concerning properties of matrices and tensors on supervector space which
feature in the main text, appear as  supertranspose,  superdeterminant, supertrace, etc \cite{D}.

\smallskip

We consider the standard basis for the
supervector spaces so that in writing the basis as a column
matrix, we first present the bosonic base, then the fermionic
one. The transformation of standard basis and its dual basis can
be written as follows:
\begin{equation}
{e'}_i=(-1)^j{K_i}\;^j e_j ,\hspace{10mm}{e'}^i={{K^{-st}}^i}_j\; e^j,
\end{equation}
where the transformation matrix $K$ has the
following block diagonal representation \cite{D}
\begin{equation}\label{A.2}
K=\left( \begin{tabular}{c|c}
                 A & C \\ \hline
                 D & B \\
                 \end{tabular} \right),
\end{equation}
where $A,B$ and $C$ are real submatrices and $D$ is pure
imaginary submatrix\footnote{For further details, one may refer
to DeWitt's book \cite{D}, p.24.}. Here we consider the matrix and
tensors having a form with all upper and lower indices written in
the right hand side.\\
The transformation properties of upper and lower
right indices to the left one for general tensors are as follows:
\begin{equation}
^iT_{jl...}^{\;k}=T_{jl...}^{ik},\qquad
_jT^{ik}_{l...}=(-1)^j\;T_{jl...}^{ik}.
\end{equation}
Let $K, L, M$ and $N$ be the matrices whose their elements indices have different positions. Then,
we define the {\it supertranspose } for these matrices as follows:
$$
{{K}^{st\;i}}_j=(-1)^{ij}\;{K_j}^{\;i},\qquad
{L^{st}_{\;i}}^{\;j}=(-1)^{ij}\;{L^j}_{\;i},
$$
\begin{equation}
M^{st}_{\;ij}=(-1)^{ij}\;M_{\;ji},\qquad
N^{st\;ij}=(-1)^{ij}\;N^{\;ji}.
\end{equation}
For the matrix  $K$  whose elements $_iK^j$  have the left index in the lower
position and the right index in the upper position, we define  the {\it supertrace}
as follows:
\begin{equation}
str K\;=\; (-1)^i \;_iK^i = {K_i}^{\;i},
\end{equation}
when $K$ is expressed in the block form  \eqref{A.2} the supertrace become
\begin{equation}
str K\;=\; tr A- trB,
\end{equation}
where 'tr' denotes the ordinary trace. \\
If the submatrix $B$ in the block form  \eqref{A.2}  is a nonsingular, then the {\it superdeterminant}
for the matrix $K$ is defined by
\begin{equation}
sdet\left( \begin{tabular}{c|c}
                 A & C \\ \hline
                 D & B \\
                 \end{tabular} \right)=det{(A-CB^{-1}D)}(det B)^{-1},
\end{equation}
and if the submatrix $A$ is nonsingular, then
\begin{equation}
sdet\left( \begin{tabular}{c|c}
                 A & C \\ \hline
                 D & B \\
                 \end{tabular} \right)=(det{(B-DA^{-1}C)})^{-1}\;(det A).
\end{equation}
If both $A$ and $B$ are nonsingular, then the {\it inverse}
matrix for \eqref{A.2} has the following form:
\begin{equation}
{\footnotesize \left( \begin{tabular}{c|c}
                 A & C \\ \hline
                 D & B \\
                 \end{tabular} \right)^{-1}=\left( \begin{tabular}{c|c}
                 $(1_m-A^{-1}C
                  B^{-1}D)^{-1}A^{-1}$&
                  $-(1_m-A^{-1}CB^{-1}D)^{-1}A^{-1}CB^{-1}$  \\ \hline

                 $-(1_n-B^{-1}DA^{-1}C)^{-1}B^{-1}DA^{-1}$  & $(1_n-B^{-1}DA^{-1}C)^{-1}B^{-1}$
                \end{tabular} \right),}
\end{equation}
where  $m$ and $n$ are dimensions of
submatrices $A$ and $B$, respectively.

\smallskip
If $f$ be a differentiable function on ${\mathbf{R}}_c^m \times {\mathbf{R}}_a^n$ (${\mathbf{R}}_c^m$ are subset of all real numbers with dimension $m$ and ${\mathbf{R}}_a^n$ are subset of all
odd Grassmann variables with dimension $n$), then relation between the left partial differentiation
and right ones is given by
\begin{equation}
\frac{\overrightarrow{\partial}}{{\partial} x^i} f \;=\; (-1)^{i(|f|+1)}\; f \frac{\overleftarrow{\partial}}{{\partial} x^i},
\end{equation}
where $|f|$ indicates the grading of $f$.\\
If $f$ be a scalar field, $\overrightarrow{\mathbf{X}}\;=\;X^i \frac{\overrightarrow{\partial}}{{\partial} x^i}$
a contravariant vector field and $\mathbf{\omega}\;=\; \omega_i dx^i$ a covariant vector field, then one finds {\it covariant derivative } in explicit components form as follows:
\begin{align}
f {\overleftarrow{\nabla}}_i &=\  (-1)^{i|f|}\; \overrightarrow{\nabla}_i f = f \frac{\overleftarrow{\partial}}{{\partial} x^i},\\[2mm]
X^i {\overleftarrow{\nabla}}_j &=\  (-1)^{j(|X|+i)}\; \overrightarrow{\nabla}_j X^i = X^i \frac{\overleftarrow{\partial}}{{\partial} x^j}+(-1)^{k(i+1)} X^k {\Gamma}^i_{\;~kj},\\[2mm]
\omega_i {\overleftarrow{\nabla}}_j &=\  (-1)^{j(|\omega|+i)}\; \overrightarrow{\nabla}_j \omega_i = \omega_i \frac{\overleftarrow{\partial}}{{\partial} x^j}- \omega_k {\Gamma}^k_{\;~ij},
\end{align}
where ${\Gamma}^i_{\;~jk}$ are called the components of the connection $\bf{\nabla}$.\\
If the supersymmetric matrix ${_AG}_B$ (its inverse denotes to
${^AG}^B$ and ${G}^{AB} = (-1)^{AB} {G}^{BA}$) be the components
of metric tensor field on a Reimannian supermanifold, then, in a
coordinate basis, the components of the {\it connection} and {\it
Reimann tensor field} are given by
\begin{align}
{\Gamma}^M_{\;~NP}\;=\; (-1)^Q\;G^{MQ} {\Gamma}_{QNP} &=\  \frac{(-1)^Q}{2}G^{MQ}
\Big[G_{QN}\frac{\overleftarrow{\partial}}{\partial
x^P}+(-1)^{NP}G_{QP}\frac{\overleftarrow{\partial}}{\partial
x^N} \nonumber \\[2mm]
&-\ (-1)^{Q(N+P)}G_{NP}\frac{\overleftarrow{\partial}}{\partial
x^Q}\Big],
\end{align}
\vspace{-6mm}
\begin{align}
R^I_{\;~JKL}&=\ -{\Gamma}^I_{\;~JK}\frac{\overleftarrow{\partial}}{\partial
x^L}+(-1)^{KL}{\Gamma}^I_{\;~JL}\frac{\overleftarrow{\partial}}{\partial
x^K}+(-1)^{K(J+M)}{\Gamma}^I_{\;~MK}{\Gamma}^M_{\;~JL} \nonumber \\[2mm]
&-\ (-1)^{L(J+K+M)}{\Gamma}^I_{\;~ML}{\Gamma}^M_{\;~JK},
\end{align}
also, for the {\it curvature tensor field}, the {\it Ricci
tensor} and the {\it curvature scalar field} we have
\begin{align}
R_{IJKL}&=\  G_{IM} R^M_{\;~JKL},\\[2mm]
R_{IJ}&=\ (-1)^{K(I+1)}\;R^K_{\;~IKJ}\\[2mm]
R &=\ {R_M}^{\;M}\;=\; {\bf str}(R_{MN}G^{NM}).
\end{align}
To lower and raise indices
denoting tensor field components, one can use of the  tensor fields $G$ and $G^{-1}$ as follows:
\begin{align}
{{T_{A_1,\cdots, A_r}}^C}_ {B_1,\cdots, B_S}\;=\;(-1)^{(E+C)(B_1,\cdots, B_S)} T_{A_1,\cdots, A_r, E, B_1,\cdots, B_S}
G^{EC}.
\end{align}

\bigskip


\begin{thebibliography}{99}
\addtolength{\itemsep}{-3mm}

\bibitem {Henneaun} M. Henneaux, L. Mezincescu,
{\it A $\sigma$-model interpretation of
Green-Schwarz covariant superstring action},
\plb{152} {1985} {340-342}.\\


\bibitem {Tseytlin} R. R. Metsaev, A. A. Tseytlin,
{\it Type IIB superstring action in $AdS_5 \otimes S^5$
background},
\npb{533} {1998} {109} [\hepth{9805028}].\\


\bibitem {Berkovits}  N.Berkovits, C.Vafa, E.Witten,
{\it Conformal Field Theory of Ads Backgrounds with
Ramond-Ramond Flux},
\jhep{03}{1999}{018} [\hepth{9902098}].\\


\bibitem{Bershadsky} N. Berkovits, M. Bershadsky, T. Hauer, S. Zhukov and B. Zwiebach,
{\it Superstring Theory on
$AdS_2 \otimes S^2$ as a Coset Supermanifold},
\npb{567} {2000} {61-86} [\hepth{9907200}].\\


\bibitem{Sorokin} D. Sorokin, A. Tseytlin, L. Wulff  and K. Zarembo,
{\it Superstring in $AdS_2 \otimes S^2 \otimes
T^6$},
[\arXivid{1104.1793}].\\


\bibitem{Busher} T. H. Buscher,
{\it Path-integral derivation of Quantum Duality in nonlinear
sigma-models},
\plb{201} {1988} {466-472}.\\


\bibitem{K.S1}  C. Klim\v {c}ik and P. \v {S}evera,
{\it Dual non-Abelian duality and
the Drinfeld double},
\plb{351} {1995} {455-462} [\hepth{9502122}].\\


C. Klim\v {c}ik and P. Severa,
{\it Poisson-Lie T-duality and loop
groups of drinfeld doubles},
\plb{372} {1996} {65} [\hepth{9512040}].\\


\bibitem {ER}  A. Eghbali, A. Rezaei-Aghdam,
{\it Poisson-Lie T-dual sigma models on supermanifolds},
\jhep{09}{2009}{094} [\arXivid{0901.1592}].\\



\bibitem{D}  B. DeWitt,
{\it Supermanifolds},  Cambridge University Press
1992.\\

\bibitem {An} Andruskiewitsch, N., {\it Lie superbialgebras and
Poisson-Lie supergroups}, Abh. Math. Semin. Univ. Hambg. 63, 147 {1993}.\\


\bibitem {ER1}  A. Eghbali, A. Rezaei-Aghdam and  F. Heidarpour,
{\it Classification of two and three dimensional Lie
super-bialgebras},
\jmp{51} {2010} {073503} [\arXivid{0901.4471}].\\


\bibitem {ER4}  A. Eghbali, A. Rezaei-Aghdam and F. Heidarpour,
{\it Classification of four and six dimensional Drinfel'd
superdoubles},
\jmp{51} {2010} {103503} [\arXivid{0911.1760}].\\


\bibitem {Tyurin} E. Tyurin and R. von Unge, {\it Poisson-Lie T-Duality: the Path-Integral Derivation},
\plb{382} {1996} {233} [\hepth{9512025}].\\


\bibitem{Sfetsos}  K. Sfetsos,
{\it Canonical equivalence of non-isometric sigma models and
Poisson-Lie T-duality},
\npb{517} {1998} {549-566} [\hepth{9710163}].\\


\bibitem {Von Unge}  R. Von Unge, {\it Poisson-Lie T-plurality}, \jhep{07}{2002}{014} [\hepth{0205245}].\\


\bibitem {N.M} A. Bossard and N. Mohammedi, {\it Poisson-Lie Duality in the String Effective Action},
\npb{619} {2001} {128-154} [\hepth{0106211}].\\


\bibitem {B}  N. Backhouse,
{\it A classification of four-dimensional Lie
superalgebras},
\jmp{19}  {1978} {2400-2402}.\\


\bibitem {ER5}  A. Eghbali and A. Rezaei-Aghdam, {\it 3 + 1 dimensional string cosmological models on the Lie
supergroup $OSP(1|2)$}, work in progress.\\



\end{thebibliography}
\end{document}